\documentclass[twocolumn]{aastex631}

\received{April 28, 2021}
\revised{July 13, 2021}
\accepted{July 20, 2021}

\def\OIII{[\mbox{O\,{\sc iii}}]$\lambda 5007$}
\def\NII{[\mbox{N\,{\sc ii}}]$\lambda 6583$}
\def\SII{[\mbox{S\,{\sc ii}}]$\lambda 6717,6731$}

\def\Ha{{H$\alpha$}}
\def\Hb{{H$\beta$}}

\def\NIIHa{[\mbox{N\,{\sc ii}}]$\lambda 6583$/H$\alpha$}
\def\OIIIHb{[\mbox{O\,{\sc iii}}]$\lambda 5007$/H$\beta$}
\def\SIIHa{[\mbox{S\,{\sc ii}}]$\lambda 6717, 6731$/H$\alpha$}

\def\kms{${\rm km}~{\rm s}^{-1}$}
\newcommand{\ergcms}	{\ifmmode {\rm erg\,cm}^{-2}\,{\rm s}^{-1} \else erg\,cm$^{-2}$\,s$^{-1}$\fi}

\shorttitle{Compact Elliptical Galaxies Hosting Active Galactic Nuclei in Isolated Environment}
\shortauthors{Rey et al.}
\DeclareUnicodeCharacter{2212}{-}
\begin{document}

\title{Compact Elliptical Galaxies Hosting Active Galactic Nuclei in Isolated Environment}

\correspondingauthor{Soo-Chang Rey}
\email{screy@cnu.ac.kr}

\author[0000-0002-0041-6490]{Soo-Chang Rey}
\affiliation{Department of Astronomy and Space Science, Chungnam National University, Daejeon 34134, Republic of Korea; screy@cnu.ac.kr}

\author[0000-0002-5037-951X]{Kyuseok Oh}
\affiliation{Korea Astronomy and Space Science Institute, Daedeokdae-ro 776, Yuseong-gu, Daejeon 34055, Republic of Korea; oh@kasi.re.kr}

\author[0000-0002-3738-885X]{Suk Kim}
\affiliation{Department of Astronomy and Space Science, Chungnam National University, Daejeon 34134, Republic of Korea}



\begin{abstract} 
We present the discovery of rare active galactic nuclei (AGNs) in nearby ($z < 0.05$) compact elliptical galaxies (cEs) located in isolated environments. Using spectroscopic data from the Sloan Digital Sky Survey (SDSS) Data Release 12, four AGNs were identified based on the optical emission-line diagnostic diagram. SDSS optical spectra of AGNs show the presence of distinct narrow-line emissions. Utilizing the black hole (BH) mass-stellar velocity dispersion scaling relation and the correlation between the narrow $L({\rm [OIII]})$/$L({\rm H\beta})$ line ratio and the width of the broad \Ha\ emission line, we estimated the BH masses of the cEs to be in the range of $7\times10^{5}$ -- $8\times10^{7}$ \(M_\odot\). The observed surface brightness profiles of the cEs were fitted with a double S\'{e}rsic function using the Dark Energy Camera Legacy Survey $r$-band imaging data. Assuming the inner component as the bulge, the $K$-band bulge luminosity was also estimated from the corresponding Two Micron All Sky Survey images. We found that our cEs follow the observed BH mass-stellar velocity dispersion and BH mass-bulge luminosity scaling relations, albeit there was a large uncertainty in the derived BH mass of one cE. 
In view of the observational properties of BHs and those of the stellar populations of cEs, we discuss the proposition that cEs in isolated environments are bona fide low-mass early-type galaxies (i.e., a nature origin).
\end{abstract}



\section{Introduction} \label{sec:intro}

In the nearby universe, compact elliptical galaxies (cEs) are a rare class of non-star-forming objects that are typically characterized by compact sizes (100 $-$ 900 pc), low stellar masses (10$^{8} - 10^{10}$ \(M_\odot\)), and very high stellar densities. Only a few hundred cEs are currently known 
\citep{Chilingarian2009, Norris2014, Chilingarian2015,Kim2020}. They conform to the low-mass regime of the mass-size relation defined by massive early-type galaxies. Although the origin of cEs is still a subject of debate \citep{Bekki2001,Martinovic2017,Du2019,UrrutiaZapata2019}, cEs are thought to be a mixture of objects formed via two main channels (i.e., nature or  nurture) depending on their local environment \citep[i.e., without or with a nearby massive host galaxy;][]{Ferre2018,Ferre2021,Kim2020}. 

The prevailing scenario is that progenitors of cEs are larger, more massive galaxies, and their outer envelopes are stripped by dissipative tidal interactions with a more massive neighboring host galaxy. In this process, cEs are the tidally stripped central remnants of their massive progenitors (i.e., stripped cEs). This scenario is supported by the fact that a large fraction of cEs are associated with an adjacent massive host galaxy \citep{Norris2014,Chilingarian2015,Janz2016,Kim2020}. In addition, the metallicities of most cEs with a massive host galaxy are higher than those of low-mass galaxies with similar masses, but are comparable to those of more massive galaxies \citep[e.g.,][]{Kim2020}. The discovery of tidal streams around a few cEs that are being stripping by their host galaxy provides further direct evidence for this scenario \citep[e.g.,][]{Huxor2011,PaudelRee2014,Ferre2018}.

However, the discovery of some isolated cEs with no neighboring massive host galaxy suggests an additional origin of cEs \citep{Huxor2013,Paudel2014}. Such cEs are intrinsically low-mass, compact galaxies formed at very early cosmic times without acquiring ex-situ stellar mass in their evolutionary histories (i.e., intrinsic cEs). In this case, they represent the low-luminosity extension of luminous elliptical galaxies, which is supported by the fact that cEs follow the scaling relations of classical elliptical galaxies 
\citep[e.g.,][]{Wirth1984,Kormendy2009,Kormendy2012}. For example, most isolated cEs follow the mass-metallicity relation of massive early-type galaxies at the low-mass regime \citep[e.g.,][]{Kim2020}. 

There is a correlation between the masses of black holes (BHs) and bulge masses of the galaxies that host them \citep[e.g.,][]{KormendyHo2013}. In this regard, predictions for different BH masses of cEs are expected, depending on the formation channel. If cEs are formed through a stripping process, they might host supermassive BHs appropriate for their massive progenitors. In contrast, if cEs are low-mass versions of classical early-type galaxies, they will instead host intermediate BHs in accordance with their low stellar masses. Moreover, the incidence of active galactic nucleus (AGN) activity also provides an important constraint on the star-formation history and environment of cEs, which are closely related to the gas reservoir that can trigger the nuclear activity within galaxies \citep[][and references therein]{Aird2019,Man2019}. Thus, the BH properties
of cEs would further provide independent tracers for probing the origins of cEs.

However, observational hints of the existence of BHs in cEs have been reported for only a handful to date \citep[e.g.,][]{Kormendy1997,VanDerMarel1997,Mieske2013,Paudel2016}. Moreover, a direct optical signature of an accreting BH has been discovered for only one cE using the Sloan Digital Sky Survey (SDSS) spectra  \citep{Paudel2016}. In this work, based on our large sample of cEs \citep{Kim2020}, we report the discovery of rare cEs that exhibit AGN activity possessing narrow-line emissions in their SDSS optical spectra. The remainder of this paper is organized as follows. In Section 2, we describe the identification of cEs with AGNs (hereafter ${\rm cE}_{\rm AGNs}$) and their structural properties. In Section 3, we present the environment of these ${\rm cE}_{\rm AGNs}$ and derive estimations of their BH masses. In Section 4, we discuss a possible formation scenario for ${\rm cE}_{\rm AGNs}$ in isolated environments. Throughout this study, we assume the cosmological parameters to be $\Omega_{m} = 0.3$, $\Omega_{\Lambda} = 0.7$, and $h_{0} = 0.73$.

\section{Sample and Analysis} \label{sec:style}

We have recently constructed a new catalog of cEs in the local volume ($z < 0.05$) using the SDSS DR12 \citep{Kim2020}. Following the conventional selection criteria of cEs, 138 cEs with sizes smaller than 600 pc were selected for the final sample \bibpunct[ ]{(}{)}{;}{a}{}{;}\citep[see][for more details on the selection of cEs]{Kim2020}. \citet{Kim2020} also investigated the stellar population properties (i.e., age and [Z/H]) of cEs based on the Lick indices obtained from the SDSS spectra by comparing them to simple stellar population model grids.

\begin{figure*}
\centering
	\includegraphics[width=0.9\linewidth]{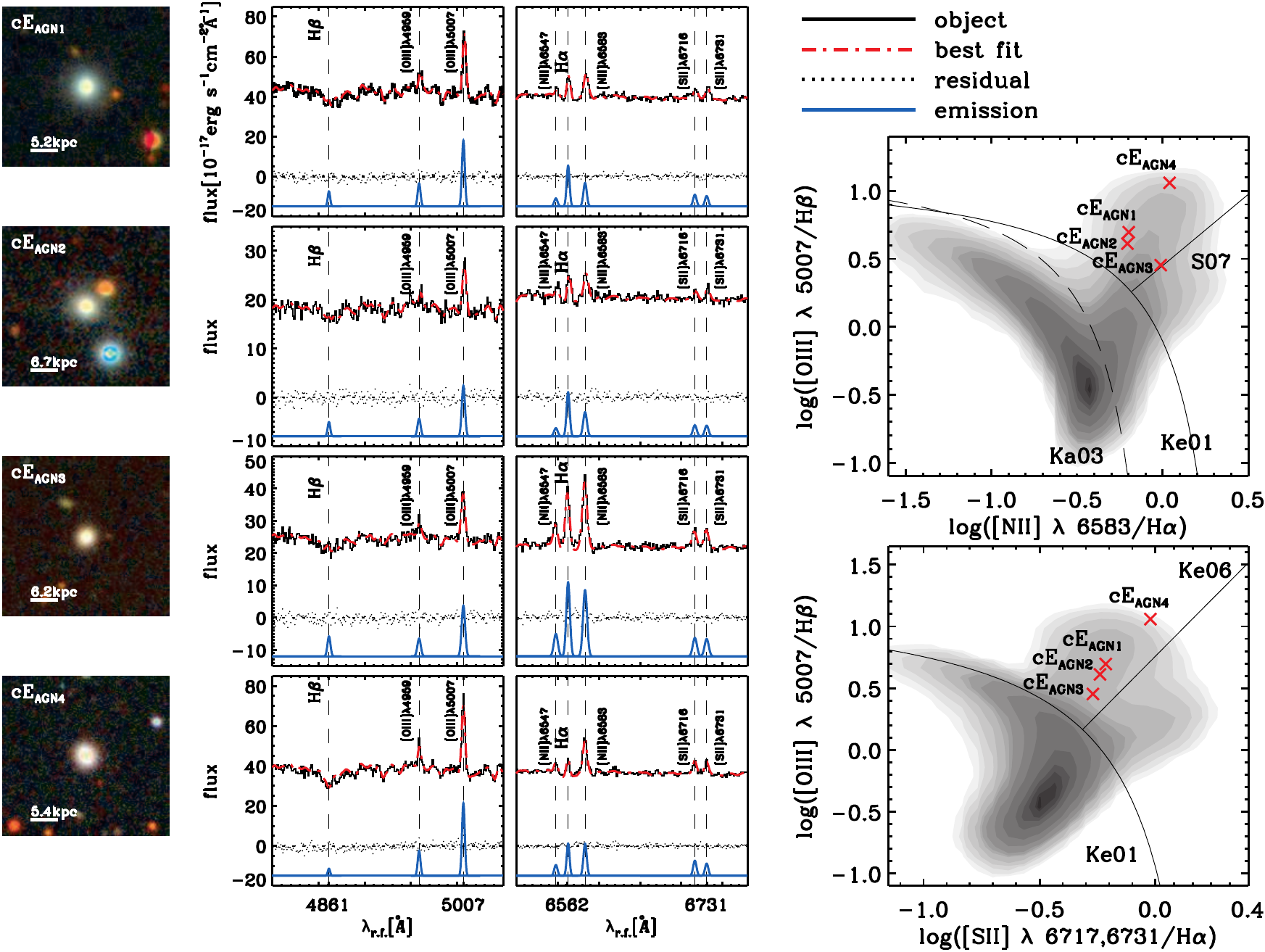} 
    \caption{$Left:$ DECaLS $g$, $r$, and $z$ composite color images of the four ${\rm cE}_{\rm AGNs}$. $Middle:$ Observed spectrum (black), the best fit (red dot-dashed), residuals (black dots), and detected emission lines (blue) of the ${\rm cE}_{\rm AGNs}$. $Right:$ BPT diagnostic diagrams. The ${\rm cE}_{\rm AGNs}$ (red crosses) and emission-line galaxies with A/N cut ($>3$ for \NII, \Ha, \OIII, \Hb, and \SII; $N \approx 180,000$) chosen from the entire OSSY catalog at $z<0.2$ \citep[filled contours;][]{Oh11} are shown.}
    \label{fig:figure1}
\end{figure*}

\subsection{Identifying Active Galactic Nuclei}

We performed spectral line fitting for the 138 cEs using the {\tt gandalf} \citep{Sarzi06}, which was later modified to measure nebular emission lines, stellar absorption lines, and stellar velocity dispersions for more than 660,000 SDSS galaxies at $z<0.2$ \bibpunct[ ]{(}{)}{;}{a}{}{,}\citep{Oh11, Oh15}. 

Our analysis of the cE SDSS spectra consisted of three major steps. First, we deredshifted the spectra and corrected the effect of Galactic foreground extinction \citep{Schlafly11}. Second, we extracted the stellar kinematics by matching the spectra with a set of stellar templates. The spectral regions that were potentially affected by nebular emission lines were masked in this process. We employed the Penalized PiXel-Fitting method ({\tt pPXF}, \citealt{Cappellari04}) using the synthesized stellar population models \citep{Bruzual03} and the MILES empirical stellar libraries \citep{Sanchez06}. Finally, we simultaneously measured the strength of the emission lines using both the stellar templates that were derived in the previous step and a set of Gaussian templates that represent the emission lines. In this process, we combined the stellar templates with a number of Gaussian emission line templates described either by single Gaussians or by sets of Gaussians that characterize doublets (e.g., middle panels in Fig.~\ref{fig:figure1}).

Using the measured emission line ratios (\NIIHa, \OIIIHb, and \SIIHa), we identified four type 2 (obscured) AGNs that possess optical spectral energy distributions dominated by stellar continua, but also narrow emission line ratios falling into the Seyfert regime in the BPT diagram  (\citealt{Baldwin81}, see right panels in Fig.~\ref{fig:figure1}). For the latter, we used the theoretical maximum starburst model \citep{Kewley01} and the empirical star formation curve \citep{Kauffmann03}. Additionally, in order to discriminate low-ionization nuclear emission line regions from Seyfert AGNs, we used the demarcation lines derived by \citet{Schawinski07} and \citet{Kewley06}. A Gaussian amplitude-over-noise ratio (${\rm A/N}$) greater than 3 was used for the emission lines (\NII, \Ha, \OIII, \Hb, and \SII). Moreover, the four ${\rm cE}_{\rm AGNs}$ of this study have been also identified by \citet{ferremateu21}.

\subsection{Structural Parameters}
We used the Dark Energy Camera Legacy Survey (DECaLS) imaging data to derive the structural parameters of the four ${\rm cE}_{\rm AGNs}$. With an average seeing of 1.18 arcsec and a pixel scale of ~0.262 arcsec/pix \citep{Dey2019}, the DECaLS $r$-band imaging of the ${\rm cE}_{\rm AGNs}$ provide better quality data than the SDSS images (see Fig.~\ref{fig:figure1}). All of the objects in each DECaLS image detected by SExtractor \citep{Bertin1996}, except the ${\rm cE}_{\rm AGN}$ itself, were removed by replacing their flux with the median background flux calculated from all the surrounding pixels in the image. 

We determined the surface brightness profiles of the ${\rm cE}_{\rm AGNs}$ using the IRAF $ellipse$ task, fitting them with a single S\'{e}rsic function and a double S\'{e}rsic function. In the case of the double S\'{e}rsic function, we fixed the outer component as an exponential profile with a S\'{e}rsic index of n $=$ 1. Based on the residuals between the observed surface brightness profiles and their model fits, we found that all four ${\rm cE}_{\rm AGNs}$ are better fitted by a double S\'{e}rsic function with S\'{e}rsic indices for the inner component in the range of 0.8 $-$ 1.1. We also derived the near-infrared $K$-band total luminosity enclosed within a Petrosian aperture by using SExtractor to obtain the photometry in the 2MASS archival images. Assuming the inner component as the bulge, the $K$-band bulge luminosity was calculated using the flux ratio between the inner and outer components \citep{Paudel2016}.  The basic properties of the four ${\rm cE}_{\rm AGNs}$ are listed in Table~\ref{tab:table1}.

\begin{deluxetable*}{ccccccccccccc}
\tablenum{1}
\small\addtolength{\tabcolsep}{-3.5pt}
\tablecaption{Basic Properties of cEs with AGN\label{tab:table1}}
\tablewidth{750pt}
\tablehead{
\colhead{Galaxy} & \colhead{R.A.} & \colhead{Decl.} & \colhead{$z$} &
\colhead{$M_{\rm r}$} & \colhead{$M_{\rm K,bulge}$}  & \colhead{Age} & \colhead{[Z/H]} &\colhead{$\sigma_{\rm *}$} & \colhead{$M_{\rm BH,\sigma}$} & \colhead{$M_{\rm BH,OSSY}$} & \colhead{$M_{\rm BH,RCSED}$} & \colhead{$M_{\rm *}$}\\
\nocolhead{} & \colhead{[deg]} & \colhead{[deg]} & \nocolhead{} &
\colhead{[mag]} & \colhead{[mag]} & \colhead{[Gyr]} & \nocolhead{} & \colhead{[\kms]} & \colhead{[log($M_\odot$)]} & \colhead{[log($M_\odot$)]} & \colhead{[log($M_\odot$)]} & \colhead{[log($M_\odot$)]}
}
\decimalcolnumbers
\startdata
${\rm cE}_{\rm AGN1}$(K12) & 55.8761  & $-$7.5854 & 0.0356 & $-$18.87 & $-$20.51  & $2.8^{+3.0}_{-1.0}$ & $-0.26^{+0.26}_{-0.27}$ & 85$\pm$6  & 6.86$\pm$0.22 & 6.59$\pm$0.08 & 6.33$\pm$0.26 & 9.23$\pm$0.02\\
${\rm cE}_{\rm AGN2}$(K28) & 144.3114 & 37.2918   & 0.0460 & $-$18.65 & $-$21.13  & $2.8^{+2.8}_{-1.0}$ & $-0.13^{+0.27}_{-0.20}$ & 91$\pm$11 & 6.99$\pm$0.27 & 6.22$\pm$0.14 & 5.92$\pm$0.37 & 9.19$\pm$0.07\\
${\rm cE}_{\rm AGN3}$(K30) & 146.2126 & 12.5123   & 0.0425 & $-$18.62 & $-$20.82  & $3.8^{+1.6}_{-2.0}$ & $-1.15^{+0.41}_{-0.38}$ & 70$\pm$11 & 6.49$\pm$0.34 & 5.87$\pm$0.09 & 5.50$\pm$0.21 & 9.03$\pm$0.02\\
${\rm cE}_{\rm AGN4}$(K41) & 159.8273 & 17.1881   & 0.0374 & $-$18.94 & $-$20.52  & $1.8^{+0.8}_{-0.6}$ & $-0.07^{+0.28}_{-0.26}$ & 75$\pm$8  & 6.63$\pm$0.27 & 7.91$\pm$0.17 & 7.23$\pm$0.27 & 9.20$\pm$0.01
\enddata
\tablecomments{(1) Galaxy name (the name of the \citet{Kim2020} catalog is in parenthesis); (2) R.A. (J2000) and (3) Decl. (J2000) from the SDSS; (4) redshift from the SDSS;  (5) DECaLS $r$-band absolute magnitude from SExtractor; (6) 2MASS $K$-band bulge absolute magnitude; (7) age and (8) metallicity of the stellar population from \citet{Kim2020}; (9) stellar velocity dispersion;  (10) BH mass derived from the BH mass-stellar velocity relation of \citet{KormendyHo2013}; (11) and (12) BH masses derived from the relation of \citet{Baron19} using the line strength measurements of \citet{Oh11} and \citet{Chilingarian17}, respectively; (13) stellar mass from \citet{Kim2020}.  
}
\end{deluxetable*}

\section{Results}
\subsection{Environment}

In our previous work, we defined the environment of our 138 cE sample \citep[see][for details]{Kim2020}. The local environment (with and without a bright host galaxy) was quantified using the projected distance from each cE to the nearest luminous ($M_{r} < -21.0$ mag) galaxy with a relative velocity of less than 500 \kms. If a cE is inside (or outside) the virial radius ($R_{vir}$) of the nearest luminous galaxy, the cE was classified as having (or not having) a host galaxy. All ${\rm cE}_{\rm AGNs}$ are located at least four times beyond the R$_{vir}$ of a luminous neighbor galaxy, indicating a low likelihood of gravitational interaction with nearby galaxies. Therefore, the ${\rm cE}_{\rm AGNs}$ are defined as those without an associated host galaxy. Following \citet{Norris2014}, the global environment (i.e., cluster, group, and field) of the ${\rm cE}_{\rm AGNs}$ was also defined using the group catalog of \citet{Tempel2014}. This resulted in all the ${\rm cE}_{\rm AGNs}$ being classified as field galaxies \citep[see][]{Kim2020}. Consequently, given the local and global environments surrounding the ${\rm cE}_{\rm AGNs}$, we suggest that all ${\rm cE}_{\rm AGNs}$ are located in isolated environments.

\subsection{Black Hole Mass and Scaling Relations}
Having measured the stellar velocity dispersions ($\sigma_{*}$), the central BH masses ($M_{\rm BH}$) of the four ${\rm cE}_{\rm AGNs}$ were estimated to be in the range of $3.1$ −- $9.8$ $\times$ $10^{6}$ \(M_\odot\) using the $M_{\rm BH}-\sigma_{*}$ relation from \citet{KormendyHo2013}. Based on the spectral resolution of SDSS, a reliable estimate for $\sigma_{*}$ is known as $\sigma_{*}>60~{\rm km}~{\rm s}^{-1}$. If the scaling relation from \citet{Gultekin09} is adopted, we obtained approximately 0.3 dex lower for the $M_{\rm BH}$.

Since a single epoch method is not applicable for our ${\rm cE}_{\rm AGNs}$ owing to a lack of broad-lines\footnote{${\rm cE}_{\rm AGN3}$ may look like a broad-line AGN, but it does not pass the cut proposed by \citet{Oh15}.}, we attempted to estimate $M_{\rm BH}$ independently using the relation between the narrow $L({\rm [OIII]})$/$L({\rm H\beta})$ line ratio and the width of the broad \Ha\ emission line (${\rm FWHM}_{\rm broad\  H\alpha}$) of \citet{Baron19}. By using our strength measurement of emission lines (\OIII\ and \Hb), the $M_{\rm BH}$ values of the ${\rm cE}_{\rm AGNs}$ were estimated to be in the range of $7\times10^{5}$ -- $8\times10^{7}$ \(M_\odot\). For comparison, we also estimated $M_{\rm BH}$ by adopting the strengths of emission lines provided by \citet{Chilingarian17}, which are approximately 0.3 dex lower. It should be mentioned that the observed relationship between the $L({\rm [OIII]})$/$L({\rm H\beta})$ and ${\rm FWHM}_{\rm broad\  H\alpha}$ of \citet{Baron19} exhibits a significant scatter in the bulk population of AGNs. As a result, regardless which line strength measurements are used, the derived $M_{\rm BH}$ of ${\rm cE}_{\rm AGN4}$ is high and has a large uncertainty. The estimated $M_{\rm BH}$ of the four ${\rm cE}_{\rm AGNs}$ and its measurement errors are listed in Table~\ref{tab:table1}. The systematic uncertainties of $M_{\rm BH}$-$\sigma_{*}$ relation and the method of \citet{Baron19} are known as $0.29$ dex \citep{KormendyHo2013} and at least $0.5$ dex \citep{Baron19}, respectively.

\begin{figure*}
\centering
	\includegraphics[width=1\linewidth]{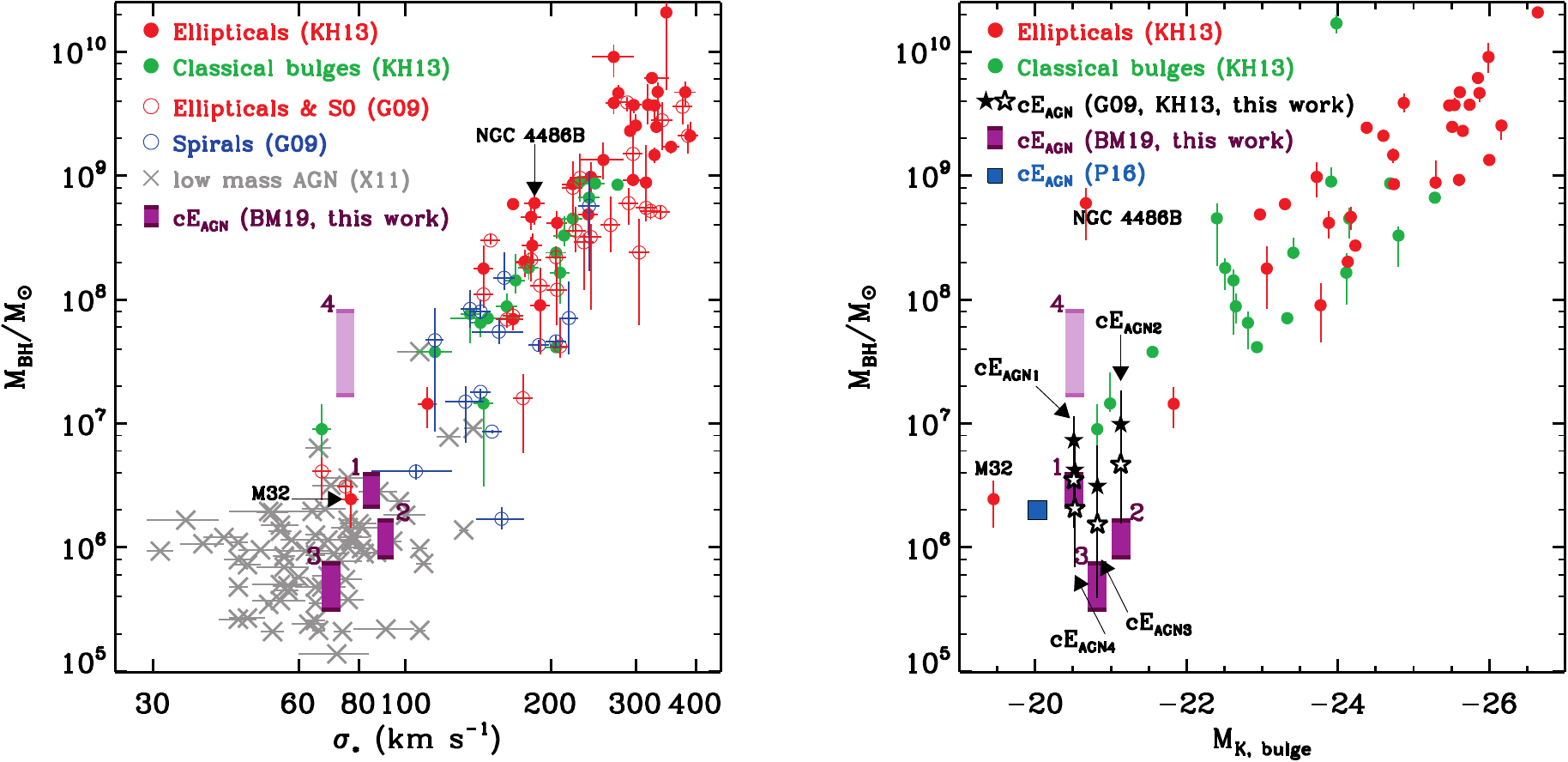} 
        \caption{$Left:$ Correlation between black hole mass ($M_{\rm BH}$) and stellar velocity dispersion ($\sigma_{*}$) of galaxies. Ellipticals (red filled circles) and classical bulges (green filled circles) from \citet{KormendyHo2013} and ellipticals/S0 (red open circles) and spirals (blue open circles) from \citet{Gultekin09} are displayed. 
        The grey crosses represent low mass AGNs \citep{Xiao11} that do not have morphology information. $Right:$ Relation between $M_{\rm BH}$ and the $K$-band absolute magnitude for ellipticals (red filled circles) and classical bulges (green filled circles) from \citet{KormendyHo2013}. The filled and open stars are the $M_{\rm BH}$ estimates of the ${\rm cE}_{\rm AGNs}$ following the scaling relations of \citet{KormendyHo2013} and \citet{Gultekin09}, respectively. In both panels, the $M_{\rm BH}$ estimates of the ${\rm cE}_{\rm AGNs}$ following \citet{Baron19} are shown with purple bars. The upper- and lower-ends of each bar are the $M_{\rm BH}$ values estimated using the line strengths of \citet{Oh11} and \citet{Chilingarian17}, respectively. The cE previously identified in isolation \citep[blue filled square;][]{Paudel2016} and two tidally stripped cEs (M32 and NGC 4486B) hosting BHs  are shown for comparison.
}
    \label{fig:figure2}
\end{figure*}

In Fig.~\ref{fig:figure2}, we present the scaling relations between the $M_{\rm BH}$ and the properties of the ${\rm cE}_{\rm AGNs}$ along with galaxies compiled from the literature. The correlations between $M_{\rm BH}$ and the stellar velocity dispersions of the galaxies (left panel) and the $K$-band absolute magnitudes of the galaxy bulges (right panel) are shown. 
The cE previously identified in isolation \citep[blue filled square; J085431.18+173730.5,][]{Paudel2016} and two tidally stripped cEs (M32 and NGC 4486B) that are known to contain BHs, are also presented in both panels. Overall, our ${\rm cE}_{\rm AGNs}$ almost overlap with the galaxies at the low-velocity dispersion and low-luminosity ends, and are not outliers from the observed scaling relations\footnote{Note that the $M_{\rm BH}$ of the ${\rm cE}_{\rm AGN4}$ derived from the method of \citet{Baron19} is highly uncertain due to a significant offset of $L{\rm ([OIII])}$/$L{\rm (H\beta)}$ from the bulk population of AGNs.}.

\begin{figure*}
\centering
	\includegraphics[width=0.8\linewidth]{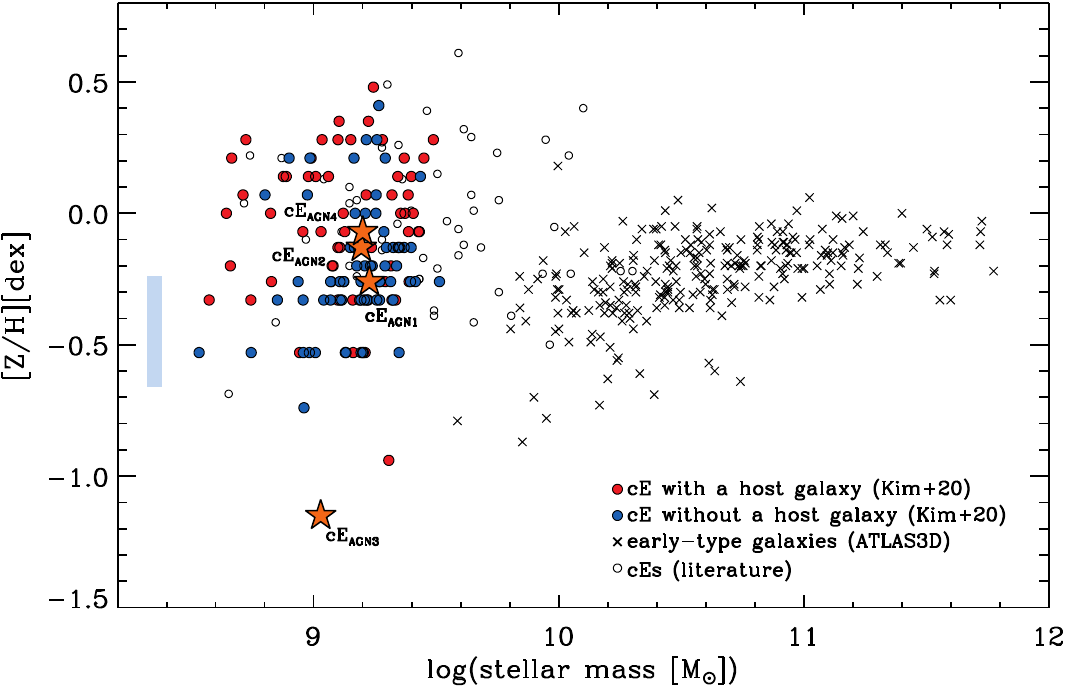} 
    \caption{$\rm[Z/H]$ vs. stellar mass distribution for cEs. The four ${\rm cE}_{\rm AGNs}$ are denoted by orange stars. The red and blue filled circles represent cEs with and without a bright host galaxy, respectively, from \citet{Kim2020}. The open circles denote cEs compiled from the literature \citep{Norris2014, Guerou15,Janz2016,Ferre2018,Ferre2021}. The crosses are early-type galaxies from ATLAS3D \citep{McDermid15}. The range of $\rm[Z/H]$ values ($-0.24$ to $-0.66$) of the ${\rm cE}_{\rm AGNs}$ provided by \citet{Chilingarian17} are denoted by the vertical bar.
}
    \label{fig:figure3}
\end{figure*}

\section{DISCUSSION AND CONCLUSION}
\subsection{Ejected Compact Ellipticals from Dense Environment?}
While the isolated cEs are considered intrinsic low-mass early-type galaxies, an alternative channel for their formation has also been suggested \citep{Chilingarian2015}. In this scenario, isolated cEs are also originally formed through tidal stripping in clusters or groups, but they are subsequently ejected from their original environments via three-body encounters.

Based on the group catalog of \citet{Saulder16}, for each ${\rm cE}_{\rm AGN}$ we searched for the nearest group. The ranges of the three dimensional and sky-projected distances of three ${\rm cE}_{\rm AGNs}$ (${\rm cE}_{\rm AGN1}$, ${\rm cE}_{\rm AGN2}$, and ${\rm cE}_{\rm AGN4}$) from their nearest groups are  4$-$6 Mpc and 2$-$6 Mpc, respectively. Considering the large physical separations, ejection from groups is improbable for these ${\rm cE}_{\rm AGNs}$. On the other hand, ${\rm cE}_{\rm AGN3}$, which is listed as a member of a very small group with only two members, is located at a distance of 5 Mpc (three-dimensional) and 0.4 Mpc (projected) from the nearest group. If the ${\rm cE}_{\rm AGN3}$ was formed through a stripping process at a very early epoch, followed by ejection from a group with a velocity comparable to the velocity dispersion of the group ($\sim$ 30~\kms), it would take longer than a Hubble time for the ${\rm cE}_{\rm AGN3}$ to move to its current location. Unless ${\rm cE}_{\rm AGN3}$ is an abnormal high-speed object on an extremely eccentric orbit \citep{Sales2007,Paudel2016}, we can rule out the possibility that ${\rm cE}_{\rm AGN3}$ is an object ejected from a group.  Moreover, it is not possible for the ejection of a cE from its host group by three-body interaction to occur concurrently with cE formation through tidal stripping by its host galaxy \citep{Chilingarian2015}.

In accordance with the mass-metallicity relation, if cEs with host galaxies are remnant cores of their massive progenitors that are formed through a stripping process, they would be more metal-rich than galaxies of comparably low-masses \bibpunct[]{(}{)}{;}{a}{}{,}\citep[][, see red circles for cEs with a host galaxy in Fig. 3]{Chilingarian2015,Janz2016,Ferre2018,Ferre2021,Kim2020}. 
In contrast, intrinsically low-mass cEs in isolated environments would follow the low-mass regime of the mass-metallicity relation \citep[][, see blue circles for cEs without a host galaxy in Fig. 3]{Kim2020}. If ${\rm cE}_{\rm AGNs}$ are stripped cEs, their massive progenitors would lose most of their original masses during the stripping process, but the metallicities of the central cores would remain almost unchanged. Therefore, they should deviate from the mass-metallicity relation of early-type galaxies, being more metal-rich than the metallicities expected for their stellar masses. 
However, as shown in Fig. 3, all the ${\rm cE}_{\rm AGNs}$ appear to fall below the observed mass-metallicity distribution of cEs with a host galaxy, conforming instead to that of cEs without a host galaxy.
This also supports the suggestion that all ${\rm cE}_{\rm AGNs}$ are intrinsic cEs formed in isolated environments rather than stripped cEs that were ejected from denser environments.

\subsection{ Origin of Isolated Compact Ellipticals: Black Hole Perspective}

The AGN fraction of galaxies is highly dependent on the environment in which AGN host galaxies reside (e.g., \citealt{Man2019} and references therein); AGNs are preferentially found in lower-density environments. Galaxies in isolated environments, on average, have higher gas reservoirs for triggering AGN activity because the environmental effects required for stripping the gas are not obvious. This is in line with that our four ${\rm cE}_{\rm AGNs}$ and the previously discovered ${\rm cE}_{\rm AGN}$  with broad emission lines of an accreting BH \citep{Paudel2016} reside in isolated environments with no nearby bright host galaxies.

Moreover, a connection between overall AGN activity and the star formations of host galaxies has been identified \citep[][ and references therein]{Aird2019}. The availability of cold gas within galaxies may drive both AGN activity and star formation. 
In this respect, the stellar populations of ${\rm cE}_{\rm AGNs}$ should also reflect star formation activity. AGNs might be rare in cEs with old ages because these galaxies are devoid of gas for feeding central BHs. It has been shown that the majority of cEs typically have stellar populations with intermediate to old ages ($>$ 8 Gyr) and solar or oversolar metallicities \citep{Ferre2018,Ferre2021,Kim2020}. 
However, a fraction of cEs also show relatively young ages ($<$ 5 Gyr) and sub-solar metallicities \citep[see Fig. 10 of][ for a dichotomy in the age distribution]{Ferre2021}. The mean ages of our ${\rm cE}_{\rm AGNs}$ are 2$-$4 Gyr, which is indicative of star formation in recent epochs. In addition, SDSS emission line diagnostics have not identified ongoing star formation. Therefore, a plausible fuel for the triggering of AGN activity in ${\rm cE}_{\rm AGNs}$ may be residual gas related to the star formation events that have recently ceased. This can provide a sustained supply of gas that could generate low-level AGN activity. 

In their recent studies, \citet{Ferre2018,Ferre2021} found that the majority of field cEs with no host galaxies exhibit slower and more extended star formation timescales than their counterparts in groups and clusters. Some isolated cEs show low star formation rates up to the present time. This type of star formation history (SFH) is typical of low-mass galaxies \citep[e.g.,][]{Thomas2005}, implying that isolated cEs are intrinsically low-mass cEs rather than objects formed from the stripping process of massive progenitors. The observed young ages and moderately low metallicities of our ${\rm cE}_{\rm AGNs}$ reinforce the hypothesis of extended SFHs of isolated cEs, which is most likely responsible for the connection with their AGN activity.

It has been proposed that cEs are a mixture of galaxies with two possible origins. Different predictions for the properties of such cEs in different parameter spaces are expected. If cEs are the tidally stripped remnants of large, more massive progenitor galaxies, the properties of the core would likely be maintained, while a large fraction of the outer envelope would be removed. However, if cEs are intrinsic low-mass objects, they would retain their characteristics and follow the low-mass end of the local scaling relations described by massive early-type galaxies. In this respect, measuring the BH mass would provide another diagnostic tool for constraining the origin of cEs \citep[see also][]{Ferre2021}. The BH masses of stripped cEs should correspond to those of massive progenitor galaxies. Thus, these cEs should deviate from the relation between BH mass and stellar mass because they have higher BH masses than galaxies with comparable stellar masses  \citep[e.g., two tidally stripped cEs, M32 and NGC 4486B, right panel of Fig. 2; see also Fig. 9 of][]{Chilingarian18}. If cEs are bona fide low-mass classical early-type galaxies, they would conform to the BH mass-stellar mass relation. All four ${\rm cE}_{\rm AGNs}$ in this study are in agreement with the observed relation between the BH masses and stellar velocity dispersions (and $K$-band luminosities) of their host galaxies. This is additional evidence that ${\rm cE}_{\rm AGNs}$ in isolated environments are intrinsic low-mass early-type galaxies revealing BH properties commensurate with their stellar masses. Accordingly, we suggest that the BH properties of cEs are one of the key parameters responsible for distinguishing between the two different cE formation processes. In combination with the results reported in this paper, by measuring BH masses for a large sample of cEs with massive nearby host galaxies, their tidally stripped origin could be further confirmed in future works.

\begin{acknowledgments}
We are grateful to the anonymous referee for helpful comments and suggestions that improved the clarity and quality of this paper. We thank Hyunjin Jeong for providing the errors of age and [Z/H] of cEs. S.C.R. and K.O. contributed equally to this work as corresponding authors. This work was supported by the National Research Foundation of Korea through grants 2018R1A2B2006445 (S.C.R.), NRF-2020R1C1C1005462 (K.O.), and NRF-2019R1I1A1A01061237 (S.K.).
\end{acknowledgments}


\bibliography{ref}{}
\bibliographystyle{aasjournal}



\end{document}